\documentstyle[twocolumn,epsfig,prb,aps,amsmath,amssymb]{revtex}
\begin{document}

\twocolumn[\hsize\textwidth\columnwidth\hsize\csname
@twocolumnfalse\endcsname

\title{
Large-N expansion based on the Hubbard-operator path integral 
representation and its application to the $t-J$ model}

\author{
Adriana Foussats and  Andr\'es Greco 
}

\address{
Facultad de Ciencias Exactas Ingenier\'{\i}a y Agrimensura 
and Instituto de F\'{\i}sica Rosario
(UNR-CONICET). 
Av.Pellegrini 250-2000 Rosario-Argentina.
}

\date{\today}
\maketitle

\begin{abstract}
In the present work 
we have developed a large-N expansion for the $t-J$ model
based on the path integral formulation for Hubbard-operators. 
Our large-N expansion formulation contains diagrammatic rules, 
in which the propagators and vertex are written 
in term of Hubbard operators.
Using our large-N formulation we have calculated, for $J=0$, 
the renormalized 
$O(1/N)$ boson propagator. We also have calculated 
the spin-spin and charge-charge
correlation functions to leading order $1/N$. 
We have compared our diagram technique and results with the existing ones 
in the literature.
\end{abstract}


\vskip2pc]
\section{Introduction}

The role of electronic correlations is an important and 
open problem in solid state physics. Its close connection with the phenomena 
of high-$T_c$ superconductivity \cite{Anderson1} makes this problem relevant 
in present days.

One of the most popular models in the context of high-$T_c$ 
superconductivity is  the $t-J$ model. 
A natural representation of the $t-J$ model is in terms of Hubbard operators
\cite{Hubbard,Izyumov}.  
Although the $t-J$ model is quadratic, 
when is written using Hubbard-operators
\cite{Izyumov},
there are several 
difficulties in the calculation of physical quantities. 
These difficulties  are mainly:
a)the complicate commutation rules of 
the Hubbard operators \cite{Hubbard} (X-operators) and, b) 
any of the model parameters can be taken to be small in a perturbative 
evaluation of the observables. 

To handle the point a), a popular method is to use  
slave particles (slave boson and fermion) 
to decouple the original $X$-operator in usual bosons and fermions
\cite{LeGuillou}.  
For point b),  one of the nonperturbative techniques (which will be 
relevant in the present paper) is the large-$N$
expansion. 
$N$ is the number of electronic degrees of freedom per site (see below)
and $1/N$ can be considered as a small parameter.  

The large-$N$ expansion was developed in the framework of slave
boson representation \cite{Kotliar,Grilli} and, using 
Baym-Kadanoff \cite{Baym} functional theory in terms of the $X$-operators
\cite{Zeyher,Zeyher1}. 

In this paper we develop  
a large-$N$ expansion for the $t-J$ model
based on the path integral representation 
for $X$-operators. 

In a series of papers \cite{Foussats,Foussats1}
we have studied the path integral formulation for Hubbard-operators.  
Our point of view is to write a path integral 
for t-J model by using
directly the Hubbard $X$-operators as fields variables, without
any decoupling. 
Our starting point was the construction of a particular 
family of first-order constrained 
Lagrangians by using the 
Faddeev-Jackiw and Dirac methods 
\cite{Faddeev}.  
We showed that 
the constrains of the model are second class. 
The canonical quantization of this constrained theory leads 
to the commutation 
rules of the Hubbard-operators.
Next, by using path-integral techniques, 
the correlation functional and effective Lagrangian were constructed.

In Ref.[\onlinecite{Foussats1}], we found a 
particular family of constrained
Lagrangians and showed that the corresponding path-integral 
can be mapped to that of the slave-boson representation in
the radial gauge \cite{Wang,Kotliar}. 
The path-integral formulation defined in [\onlinecite{Foussats1}] 
is the starting point for our large-N expansion. 

In section II 
we give a summary of the formalism   
which will be useful for understanding our large-N expansion.
In section III,
we develop our large-N method. We present explicit calculations 
for the case $J=0$ of the $t-J$ model ($U$-infinite Hubbard model).
In    section IV, we calculate spin-spin and charge-charge 
correlation functions to leading order of large-N expansion. 
Finally, in section V 
we give our main discussions and conclusions. \\

\section{Preliminaries and definitions}
\label{model}

In this section we give a summary of the main results obtained in 
[11], which are the starting point for our large-N expansion.

As we mentioned in the introduction, although the $t-J$ model is quadratic
when is written using X-operators, the complicate commutation rules 
between them make this model highly nontrivial.
To attack the problem, our method works as follows:  

a) We construct a Lagrangian written in terms of Hubbard operators

\begin{eqnarray}
L = \sum_i a_{i \alpha \beta}(X){\dot{X_i}}^{\alpha \beta} - H(X)
\end{eqnarray} 

where the $t-J$ model Hamiltonian $H(X)$, with the addition of a 
chemical potential 
$\mu$, reads as: 

\begin{eqnarray}
H(X) &=& \sum_{i,j,\sigma}\;t_{ij}\; 
X_{i}^{\sigma 0} X_{j}^{0 \sigma} 
+ \frac{1}{2} \sum_{ij;\sigma} J_{ij}(X_{i}^{\sigma {\bar{\sigma}}} 
X_{j}^{{\bar{\sigma}} \sigma} \nonumber\\ 
&-& X_{i}^{\sigma \sigma} X_{j}^{\bar{\sigma} \bar{\sigma}}) 
-\mu\sum_{i,\sigma}\;X_{i}^{\sigma \sigma}.
\end{eqnarray} 

In (1) the coefficients 
$a_{i \alpha \beta}(X)$ are unknown and must be determined.

The indices $\alpha$ and $\beta$ can take both $0$ value (empty state)
or  spin index $\sigma = \pm$ ($up$ and $down$ state, respectively). 
The five Hubbard 
${\hat X}$-operators $X^{\sigma \sigma'}$
and $X^{0 0}$ are boson-like and the four Hubbard ${\hat X}$-operators 
$X^{\sigma 0}$ and $X^{0 \sigma}$ are fermion-like \cite{Hubbard}. 
In (2), $t_{ij}$ and
$J_{ij}$ are hoping and exchange parameters, respectively,
between sites $i$ and $j$.  
 
b) We introduce (via Lagrangian multipliers $\lambda^{\alpha \beta}$)  
a set of bosonic constrains $\Omega^{\alpha \beta}$ 
which must be also determined. 

c) After these definitions and proposals, we impose that the Dirac brackets 
\cite{Faddeev}  
between X-variables obtained using our constrained theory 
verify the correct commutation rules for Hubbard operators.
With this prescription, we obtain a set of differential equations for 
the coefficients $a_{i \alpha \beta}(X)$ and constrains $\Omega_a$.
A particular solution for the coefficients is

\begin{eqnarray}
a_{i 0 \sigma} = \frac{i}{2 X_{i}^{0 0}}\; X_{i}^{ \sigma 0}\;\;,\;\;
\hspace{0.4cm} a_{i \sigma 0} = \frac{i}{2 X_{i}^{0 0}}\; X_{i}^{0 \sigma}\
\end{eqnarray}

and for the constrains, 

\begin{eqnarray}
\Omega_{i}^{0 0} = X_{i}^{0 0} + 
\sum_{\sigma} X_{i}^{\sigma \sigma} - 1 = 0\;,
\end{eqnarray}

\begin{eqnarray}
\Omega_{i}^{\sigma \sigma'} = X_{i}^{\sigma \sigma'} - \frac{X_{i}^{\sigma 0}
X_{i}^{0 \sigma'}}{X_{i}^{0 0}} = 0\;.
\end{eqnarray}

The boson-like Lagrangian coefficients are all zero. The  
constrains are second class because none of them commutes 
with all the other ones. 

Consequently,  the dynamics is given by the Lagrangian 

\begin{eqnarray}
L(X, \dot{X}) = - \frac{i}{2}
\sum_{i, \sigma}\frac{({\dot{X_{i}}}^{0 \sigma}\;X_{i}^{\sigma 0} 
+ {\dot{X_{i}}}^{\sigma 0}\;
X_{i}^{0 \sigma})}{X_{i}^{0 0}} - H(X)\;
\end{eqnarray} 

and the constrains (4) and (5).

Now, in order to write the path integral representation we use the 
method of Faddeev-Senjanovich \cite{SE}. This method is currently 
used in quantum field theory 
to obtain the path integral of constrained systems.   

The partition function reads as:

\begin{eqnarray}
Z & = & \int {\cal D}X_{i}^{\alpha \beta}\;
\delta[X_{i}^{0 0} + \sum_{\sigma} X_{i}^{\sigma \sigma}-1]\;
\delta[X_{i}^{\sigma \sigma'} - \frac{X_{i}^{\sigma 0}
X_{i}^{0 \sigma'}}{X_{i}^{0 0}}]\nonumber\\
& \times& 
\;(sdet M_{AB})_{i}^{\frac{1}{2}}
exp\;(i \int dt\;L(X, \dot{X}))\;.
\end{eqnarray}

\noindent
where the superdeterminant

\begin{eqnarray}
(sdet M_{AB})^{\frac{1}{2}} = 1/\frac{1}{(-X_{i}^{0 0})^{2}}\;.
\end{eqnarray}

is just the superdeterminant 
of the Dirac matrix 
\cite{Faddeev}  
(or equivalently the sympletic matrix of the Faddeev-Jackiw method)
formed with the set of all the second class constrains of the theory.

The power two on $X^{0 0}$ in
$(sdet M_{AB})^{\frac{1}{2}}$ appears because we are working with two spin 
projections. This point will be very important for our large-N extension 
in the next section.   

Note the first constrain (4) (the first $\delta$-function in (7)) is the 
completeness condition 
which means that "double occupancy" at each site is forbidden. 
The constrains (5) (the second $\delta$-function in (7)) 
is obtained in our formalism as a consequence of imposing that 
the commutation rules of X-operators must be fulfilled.  

From the above equations our theory corresponds to a configuration in 
which the bosons are totally constrained and the dynamics is carried out only 
by the fermions. As it was shown in Ref[11], the path integral formalism 
corresponding to this dynamical situation is mapped in the slave 
boson representation.  

Equation (7) is our main formula to develop the large-N expansion.

\section{The large-$N$ expansion}
\label{stripfil}

In this section, we present the 
large-$N$ expansion in the framework of the
path integral representation for Hubbard operators. 
We present our explicit results for the $J=0$ case 
($U$-infinite Hubbard model) which will be very useful to show 
how our method works and to compare our results with  
previously ones  
found in the literature.

Our starting point is the partition function (7) 
written in the Euclidean form: 

\begin{eqnarray}
Z&=&\int {\cal D}X_{i}^{\alpha \beta}\;
\delta[X_{i}^{0 0} + \sum_{\sigma} X_{i}^{\sigma \sigma}-1]\;
\delta[X_{i}^{\sigma \sigma'} - \frac{X_{i}^{\sigma 0}
X_{i}^{0 \sigma'}}{X_{i}^{0 0}}] \nonumber \\
&\times&(sdet M_{AB})_{i}^{\frac{1}{2}}
exp\;(- \int d\tau\;L_E(X, \dot{X}))\;.
\end{eqnarray}
 
The Euclidean Lagrangian $L_E(X,\dot{X})$ in (9) is:

\begin{eqnarray}
L_E(X, \dot{X}) =  \frac{1}{2}
\sum_{i, \sigma}\frac{({\dot{X_{i}}}^{0 \sigma}\;X_{i}^{\sigma 0} 
+ {\dot{X_{i}}}^{\sigma 0}\;
X_{i}^{0 \sigma})}
 {X_{i}^{0 0}}
+ H(X)\;.
\end{eqnarray} 

First,  we make the following steps in the path integral (9):

a) we integrate over the boson variables $X^{\sigma \sigma'}$ using
the second $\delta$-function in (9),

b) the spin index $\sigma=\pm$, is extended to a new index $p$ 
running from $1$ to $N$. In order to get a finite theory in the 
$N$-infinite limit,  we re-scale 
the hoping $t_{ij}$ to 
$t_{ij}/N$
in the Hamiltonian,  

c) the completeness condition 
($X_{i}^{0 0} + \sum_p X_{i}^{pp}=N/2$)
can be exponentiated, as usual,  
by using the Lagrangian multipliers $\lambda_i$,

d) in order to prepare the path integral for the large-N expansion,  
we write the boson fields in terms of static mean-field values and
dynamic fluctuations 

\begin{eqnarray} 
X_{i}^{0 0} = N r_{0}(1 + \delta R_{i})\;, 
\end{eqnarray} 

\begin{eqnarray}
\lambda_{i} = \lambda_{0} + \delta{\lambda_{i}}\;,
\end{eqnarray} 

e) finally,    
we make the following change of variables 

\begin{eqnarray}
f^{+}_{i p} = \frac{1}{\sqrt{N r_o}} X_{i}^{p 0},
\end{eqnarray}

\begin{eqnarray}
f_{i p} = \frac{1}{\sqrt{N r_o}} X_{i}^{0 p},
\end{eqnarray}

By following the steps a-e, we find the  
effective Lagrangian for the $J=0$ case

\begin{eqnarray}
L_{eff}& = & - \frac{1}{2}\sum_{i,p}\left(\dot{f_{i p}}f^{+}_{i p}
+ \dot{f^{+}_{i p}}f_{i p}\right) \frac{1}{(1 + \delta R_{i})} \nonumber \\
& + & \sum_{i,j,p}^{N}\;t_{ij}\ r_{o}f^{+}_{i p}f_{j p}
 -  (\mu - \lambda_{0})\;\sum_{i,p}\;f^{+}_{i p}f_{i p} 
\frac{1}{(1 + \delta R_{i})} \nonumber\\
& + &N\;r_{0}\;\sum_{i}\delta{\lambda_{i}}\;\delta R_{i} + \sum_{i,p}
f^{+}_{i p}f_{i p}\frac{1}{(1 +  \delta R_{i})}\;
\delta{\lambda_{i}}\nonumber\\
& + &L_{ghost}.
\end{eqnarray}

In (15), $L_{ghost}$ is defined via the 
exponentiation of the superdenterminant
using ghost fields. 
After the extension to large-$N$,  
the superdeterminant becomes:  

\begin{eqnarray}
(sdet M_{AB})_{i}^{\frac{1}{2}} = \frac{{(Nr_{0})}^N}{\left(\frac{-1}
{1 + \delta R_i}\right)^{N}}\;.  
\end{eqnarray}

\noindent
The numerator of (16) together with the Jacobian 
of the transformations (11-14) 
contribute to the path-integral normalization factor. 
The denominator of (16) can be seen as a ${N \times N}$ diagonal
matrix and the integral representation is given in terms of complex boson 
ghost fields ${\cal Z}_{p}$. Therefore, 

\begin{eqnarray}
L_{ghost}({\bf {\cal Z}}) = - \sum_{i p}\;
{\bf {\cal Z}}^{\dag}_{i p}\left(\frac{1}{1 + \delta R_i}\right)
{\bf {\cal Z}}_{i p}\;.
\end{eqnarray}

To implement the $1/N$ expansion, the nonpolynomial $L_{eff}$ 
should be developed in powers of $\delta R$. 
Up to order $1/N$, it is sufficient to retain terms up to $\delta R^2$.
Then, 

\begin{eqnarray}
&L_{eff}&=-\frac{1}{2}\sum_{i,p}^{N}\left(\dot{f_{i p}}f^{+}_{i p}
+ \dot{f^{+}_{i p}}f_{i p}\right) (1 - \delta R_{i} + \delta
R_{i}^{2}) \nonumber \\ 
&+&\sum_{i,j,p}^{N}\;t_{ij} r_{o} f^{+}_{i p}f_{j p} 
- \mu \;\sum_{i,p}\;f^{+}_{i p}f_{i p}  
(1 - \delta R_{i} + \delta R_{i}^{2}) \nonumber \\ 
&+&N\;r_{0}\;\sum_{i}\delta{\lambda_{i}}\;\delta R_{i}  
+\sum_{i,p}
f^{+}_{i p}f_{i p}(1 - \delta R_{i} + \delta R_{i}^{2})\;
\delta{\lambda_{i}} \nonumber \\
&-&\sum_{i p}\;
{\bf {\cal Z}}_{i p}^{\dag}\left(1- \delta R_i+ \delta R^2_i\right)
{\bf {\cal Z}}_{i p},
\end{eqnarray} 

where we have changed $\mu$ to $\mu-\lambda_0$ and dropped constant and 
linear terms in the fields.

Looking at the effective Lagrangian (18), the
Feynman rules can be obtained as usual. The bilinear 
parts give rise to the propagators and the remaining
pieces are represented by vertices.
Besides, we assume the equation (18) written in the momentum
space once the Fourier transformation was performed.

In summary, the Feynman rules are:\\ 

i) Propagators: We associate with the two component 
$\delta X^{a} = (\delta R\;,\;\delta{\lambda})$ boson field, the propagator

\begin{eqnarray}
D_{(0) ab}(q,\omega_{n})=
\left(
\begin{array}{cc}
0  & {\frac{1}{Nr_{0}}}  \\
{\frac{1}{Nr_{0}}}      & 0
\end{array}
\right)
\end{eqnarray}

\noindent
which is represented by a dashed line in Fig.1
connecting two generic components 
a and b.

The quantities $q$ and $\omega_{n}$ are the momentum
and the Bose Matsubara frequency of the bosonic field, respectively.

We associate with the N-component fermion field $f_{p}$, the propagator 

\begin{eqnarray}
G_{(0)pp'}(k, \nu_{n}) = - \frac{\delta_{pp'}}{i\nu_{n} -
(\varepsilon_{k} - \mu )}
\end{eqnarray}

\noindent
which is represented by a solid line in Fig.1 
connecting two generic components 
p and p'.

\noindent
In (20) the electron bare dispersion was defined as 
$\varepsilon_{k} = -2tr_{o} (cosk_x+cosk_y)$, where $t$ is 
the hopping between nearest neighbors sites on the square lattice.

The quantities $k$ and $\nu_{n}$ are the momentum
and the Fermionic  Matsubara frequency of the fermionic field, respectively.

We associate with the N-components ghost field ${\bf {\cal Z}}_{p'}$, 
the propagator 

\begin{eqnarray}
{\cal D}_{pp'} = - \delta_{pp'} 
\end{eqnarray}

\noindent
which is represented by a dotted line in Fig.1 
connecting two generic components 
p and p'.\\

ii)Vertices: The expressions of the different three-leg 
and four-leg vertices 
are 

\begin{eqnarray}
\Lambda^{pp'}_{a} = (-1)\; \left(\frac{i}{2}(\nu + \nu') 
+ \mu\;;\;1\right)\;\delta^{pp'}\, 
\end{eqnarray}

\noindent
representing the interaction between two 
fermions and one boson (Fig.1a);

\begin{eqnarray}
\Lambda^{pp'}_{ab} = (- 1)  
\left(
\begin{array}{cc}
- \frac{i}{2} (\nu + \nu') - \mu   &- \frac{1}{2}      \\
- \frac{1}{2}                                          &0 
\end{array}
\right)\delta^{pp'},  
\end{eqnarray}

\noindent
representing  the interaction between two 
fermions and two bosons (Fig.1b);

\begin{eqnarray}
\Gamma^{a}_{pp'} = (-1) (\delta_{pp'}\;,\;0),
\end{eqnarray}

\noindent
representing the interaction between two ghosts and one boson 
(Fig.1c); and

\begin{eqnarray}
{\Gamma}^{ab}_{pp'} = (-1)  
\left(
\begin{array}{cc}
-1  &0  \\
0  &0  
\end{array}
\right) \delta_{pp'}\, \nonumber\\ 
\end{eqnarray}

\noindent
representing the interaction between two bosons and two ghosts
(Fig.1d).
Each vertex conserves the momentum and energy.

In the $N$-infinite limit (order $O(1)$), we only have free fermions with 
renormalized band due to correlations,  
$\varepsilon_{k} = -2tr_{0} (cosk_x+cosk_y)$. For a given value of $\mu$,  
$r_{0}$ (which from (11) and the completeness condition is equal to $\delta/2$,
where $\delta$ is the hole doping away from half-filling)   
must be determined self-consistently. This result is in agreement 
with previous calculation \cite{Wang,Zeyher}.  

The bare boson propagator (19), which is of order $O(1/N)$,  
is renormalized by a series of electronic bubbles which contribute
also in $O(1/N)$. Then, in $O(1/N)$, the boson propagator is 
dressed by electronic interactions.

By looking at the diagrammatic, it can be seen that
the irreducible boson self-energy $\Pi_{ab}$ is given 
by the sum of the contributions corresponding to the one-loop 
diagrams (in order $1/N$) seen in Fig.1e.

\begin{figure}
\begin{center}
\epsfig{file=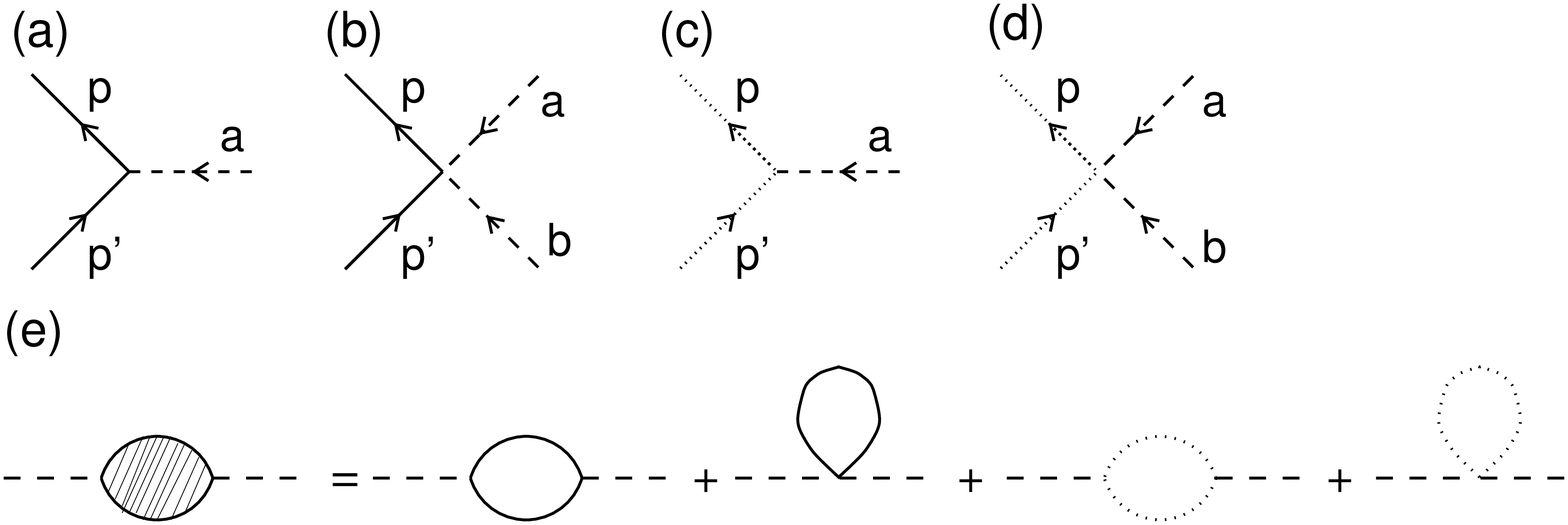,width=8.0cm,angle=0}
\end{center}
\caption{Types of vertex, up to order $O(1/N)$, between 
two 
fermions and one boson (a),
two 
fermions and two bosons (b),
two 
ghosts and one boson (c) and,
two 
bosons and two ghosts (d). (e) represents the sum of all 
one-loop diagrams contributing
to the irreducible boson-self energy. 
}
\label{}
\end{figure}

The presence of the last two diagrams in Fig.1e involving ghost fields
is very important. It is possible to show that these two diagrams 
exactly cancel the infinities,
due to the frequency 
dependence of our vertices,
of the two first diagrams in Fig.1e.
Therefore, using our Feynman rules, the final (and finite) result for each
component of the boson-self energy $\Pi_{ab}$ is:

\begin{eqnarray}
\Pi_{RR}& ( &q, \omega_{n}) = -\frac{N}{N_{s}}\;\frac{1}{4}
\sum_{k}\;\left[
2\;n_{F}(\varepsilon_{k} - \mu)(\varepsilon_{k + q} - \varepsilon_{k})
\right.\nonumber\\
& + &\left.(\varepsilon_{k + q} + \varepsilon_{k})^{2}\; 
\frac{[n_{F}(\varepsilon_{k + q} - \mu) - n_{F}(\varepsilon_{k} - \mu)]}
{- i \omega_{n} + \varepsilon_{k + q} - \varepsilon_{k}}\right]\;,
\end{eqnarray}

\begin{eqnarray}
\Pi_{\lambda R}(q, \omega_{n}) &=& - \frac{N}{N_{s} }\;\frac{1}{2}
\sum_{k}\;(\varepsilon_{k + q} + \varepsilon_{k}) \nonumber \\
&\times & \frac{[n_{F}(\varepsilon_{k + q} - \mu) - n_{F}(\varepsilon_{k} - \mu)]}
{- i \omega_{n} + \varepsilon_{k + q} - \varepsilon_{k}}  
\end{eqnarray}

\noindent
and, 

\begin{eqnarray}
\Pi_{\lambda\lambda}(q, \omega_{n}) = - \frac{N}{N_{s}}\;
\sum_{k}\;\frac{[n_{F}(\varepsilon_{k + q} - \mu) 
- n_{F}(\varepsilon_{k} - \mu)]} 
{- i \omega_{n} + \varepsilon_{k + q} - \varepsilon_{k}}\;.  
\end{eqnarray}

From the Dyson equation $(D_{ab})^{-1} = (D_{(0) ab})^{-1} - \Pi_{ab}$
the dressed  components $D_{ab}$ of the boson propagator can be found. 

Our expressions (26)-(28) agree with the previously 
irreducible boson self energies calculated in 
the context of the large-$N$ slave
boson approach \cite{Kotliar}.

The ghost fields interact only with the boson fields (Fig.1c and Fig.1d). 
The only role of the ghost fields, in order  1/N, 
is to cancel infinities in the boson self energy $\Pi_{ab}$ due to the 
frequency dependence of our vertices (22) and (23). 
For the evaluation of quantities to higher orders than 1/N,  
we must include and 
check the contribution of the ghost fields in each particular calculation.
Moreover it should be noted 
that in this case we must consider terms beyond $\delta R^2$
in developing the nonpolinomial parts of $L_{eff}$. Although this 
seems to be 
complicated, all the contributions can be controlled by the small 
number 1/N via the vertices and propagators.
At this point we want to remark that so far not many calculations go 
beyond the order 1/N. 
In order 1/N, diagramatic rules are well defined 
in the present paper.

In summary, we have developed a 
diagrammatic  technique for a large-$N$ expansion
in the same spirit of the 
large-$N$ expansion in quantum field theory. It means,
via the order of the propagators and vertices, we can determine 
the order of the diagram contribution. 
Note that our Green's functions are calculated in terms of the original  
Hubbard operators. For example, from (13) and (14) we can see that our 
fermions $f_{ip}$ are always proportional to the Fermi-like $X$-operator
$X^{op}$ and not only in the leading order like in the slave-boson formulation
\cite{Wang}.

\section{Charge-charge and spin-spin correlation function}
\label{Torder}

In this section we calculate numerically the charge-charge and spin-spin 
correlation functions on the square lattice for the nearest neighbor 
hopping $t$. We choose $t=1.0$ as the energy unit and 
the temperature $T=0.0K$. 

We can define retarded density-density $\tilde{D}$ and spin-spin $\tilde{S}$ 
Green's functions
as \cite{Gehlhoff}

\begin{eqnarray}
{\tilde{D}}_{ij}=\frac{1}{N} \sum_{pq} <T_{\tau} X^{pp}_i X^{qq}_j>
\end{eqnarray}

\noindent
and 

\begin{eqnarray}
{\tilde{S}}_{ij}=<T_{\tau} X^{pq}_i X^{qp}_j>,
\end{eqnarray}

\noindent
respectively.

Using $\sum_q X^{qq}_i=N/2- X^{00}_i$ and (11) we find for $\tilde{D}$,
in $O(1)$, in the Fourier space 

\begin{eqnarray}
\tilde{D}(Q,\omega_n)=-N {(\frac{\delta}{2})}^2 D_{RR}(Q,\omega_n)
\end{eqnarray}

Using the constrains (5), 
it is easy to prove that $\tilde{S}$, in $O(1)$, 
is the electron bubble   
formed with two bare electron Green's functions (20). 
The analytical expression 
for $\tilde{S}$ agrees with $\Pi_{\lambda \lambda}$. 
 
In Fig.2, we show $D(Q,\omega)=-Im[\tilde{D}(Q,\omega)]$ and 
$S(Q,\omega)=-Im[\tilde{S}(Q,\omega)]$ using $\eta=0.1t$ in the analytical 
continuation $i\omega_n=\omega+i\eta$.

To make explicit comparison with exact diagonalization \cite{Tohyama} 
and analytical \cite{Gehlhoff} calculation  we choose the same densities 
and momenta $Q$ of these two references. 

In Fig.2a, we plot $D(Q,\omega)$ (solid line) and $S(Q,\omega)$
(dashed line) for doping $\delta=0.12$.  
In Fig.2b, we present similar results for 
doping $\delta=0.78$.  
In agreement with analytical calculation 
based on $X$-operator canonical approach 
\cite{Gehlhoff}
(see also slave boson calculation \cite{Wang}), there is no collective 
excitations (like magnons) in the spin-spin 
correlation function to leading order of the $1/N$ expansion. The spin-spin
correlation functions is determined by the particle-hole bubble with 
the renormalized electron dispersion by correlations. 
In contrast to $S(Q,\omega)$, $D(Q,\omega)$ contains collective excitations
(zero sound) 
that can live inside or outside of the spectrum \cite{Gehlhoff}.  

For low doping (Fig.2a) the collective 
excitations lie outside the particle-hole spectrum. Therefore,
they show up as sharp peaks. In contrast, for high doping (Fig. 2b) 
the collective excitations lie inside the particle-hole spectrum
and they show up as broad peaks. The dominance of collective excitations 
in $D$, for low doping, can be clearly seen for $Q=(\pi,\pi)$.

It is also very important to point out that spin and charge 
fluctuations occur at low doping (Fig.2a), near half filling, 
on different energy scales.  
This picture disappears at high doping as it can be seen in Fig.2b. 
This behavior is not present in the uncorrelated case where spin and charge
excitations are essentially degenerate. 

The agreement between our results and previous ones (see Fig.1 of 
Ref[\onlinecite{Gehlhoff}]) 
is excelent, proving the consistency of our large-$N$ method.\\

\begin{figure}
\begin{center}
\epsfig{file=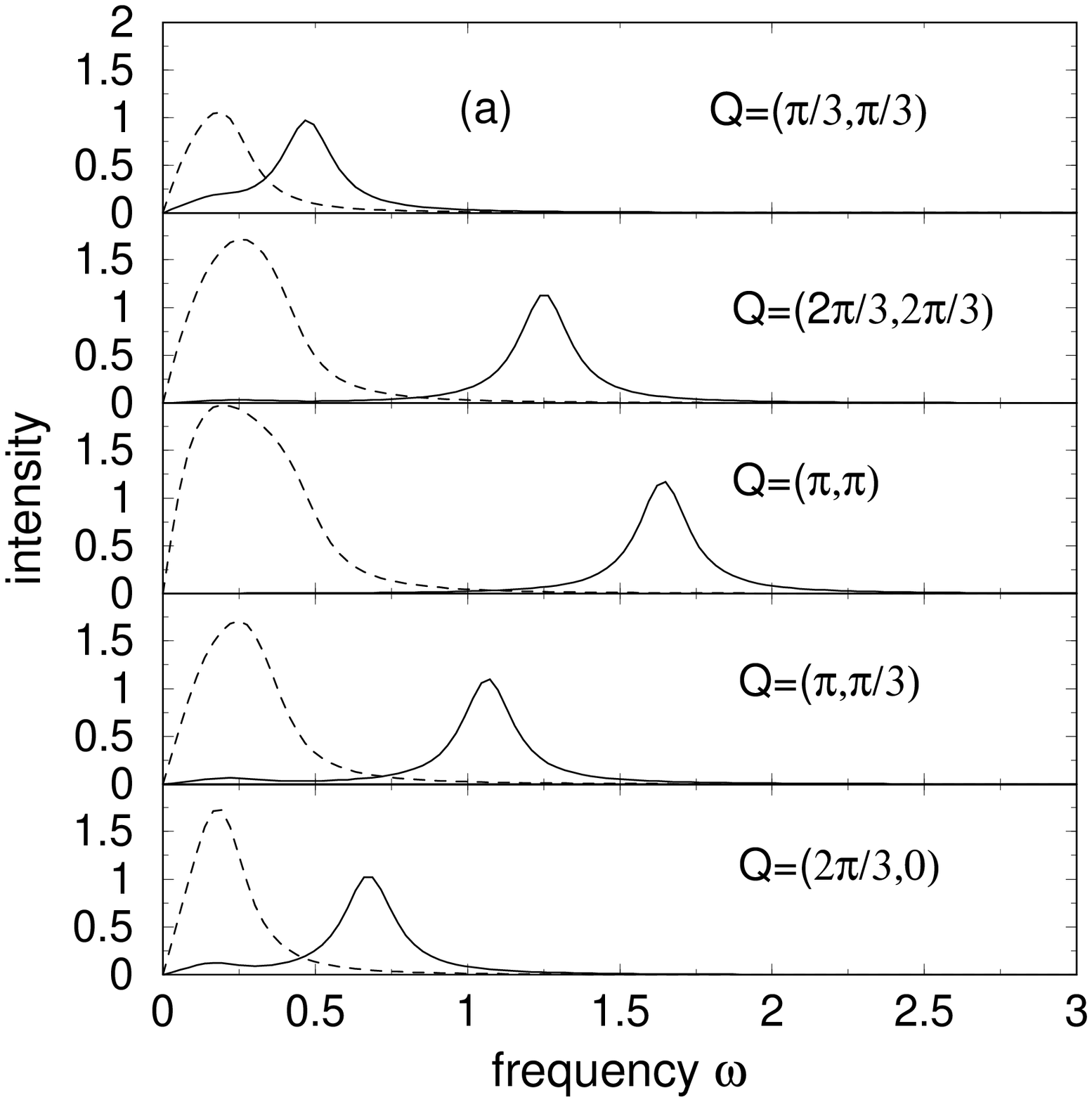,width=6.8cm,angle=-90}
\epsfig{file=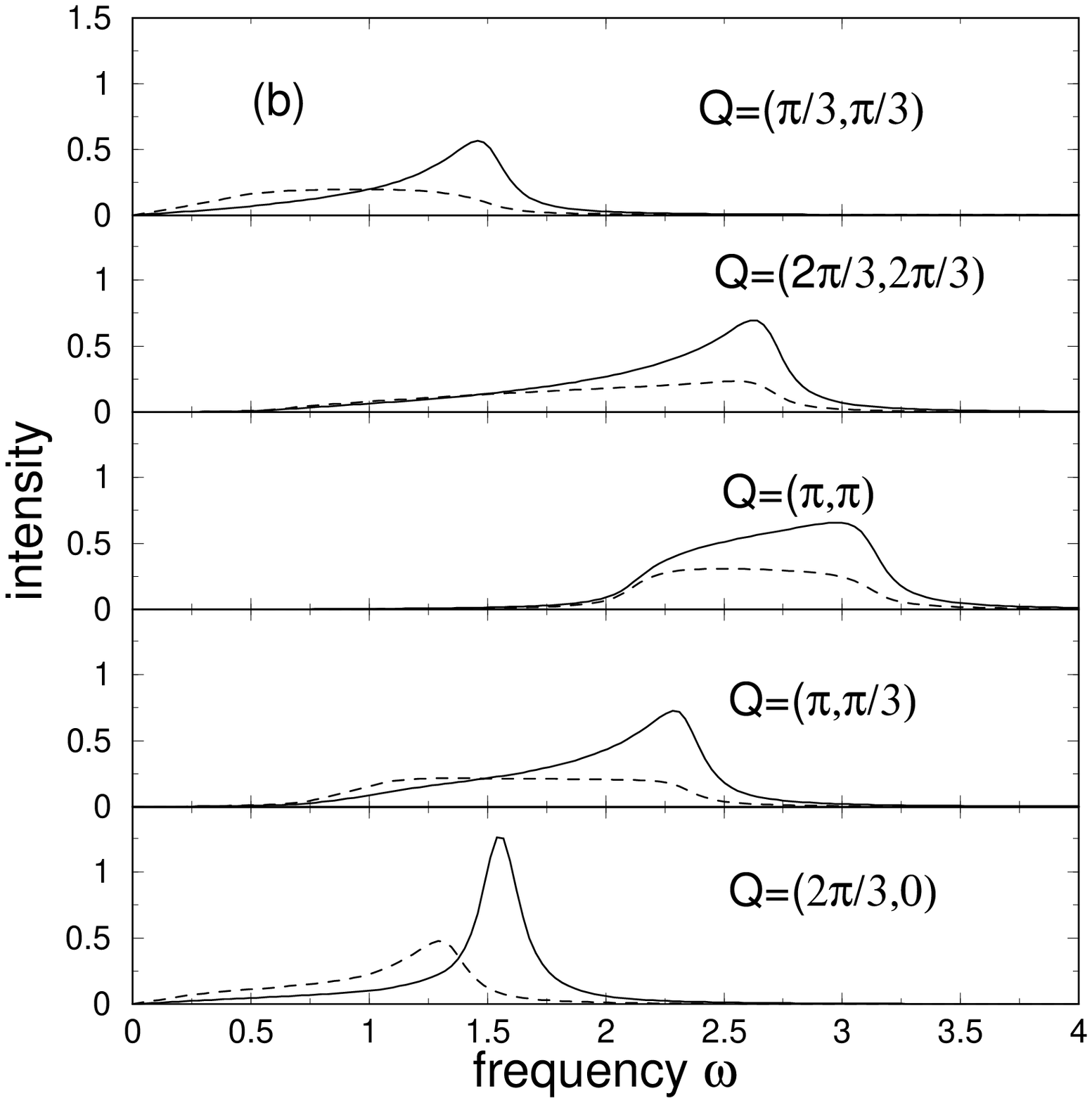,width=6.8cm,angle=-90}
\end{center}
\caption{
Spin-spin (dashed lines) and charge-charge (solid lines) correlation 
functions 
as a function of $\omega$ for several momenta for dopings $\delta=0.12$
(a) and $\delta=0.78$ (b). 
}
\label{fig1}
\end{figure}

\section{Conclusions and discussions} 

Based on our Hubbard operator path-integral representation, defined for 
the first time in Ref.[\onlinecite{Foussats1}], 
we have developed a large-$N$ expansion for the $t-J$ model. 

Because our path integral is written in terms of $X$-operators, 
we do not need any decoupling scheme for the Hubbard operators. Then, 
we do not invoke a gauge invariance or Bose condensation,  
like in the slave boson formulation \cite{Wang}.
Our dynamical variables are the original Hubbard operators 
and, there is no need to introduce spinons and holons excitations.

The use of a path integral allows us to define a Feynman diagram technique 
which is a very important and interesting 
problem in electron correlated systems
\cite{Izyumov1}. 

As we mentioned previously, the large-$N$ 
expansion was formulated using canonical
quantization in terms of $X$-operator \cite{Zeyher,Zeyher1} or 
slave boson path integral approach \cite{Wang}. 
The first one cannot be defined in a diagrammatic form   
while the second one can be developed in terms of 
Feynman diagrams but needs the introduction of spinons and holons.
In the formulation presented in this paper we have diagrammatic rules: 
propagators and vertices  in terms
of the original $X$-operators.

One important point is that in our diagram technique  the fermion operator 
$f_{\sigma}$ is proportional to the fermion $X$- 
operator $X^{0 \sigma}$ for all orders in the large $N$ expansion. 
This is very important for the interpretation of the experimental spectral
susceptibilities.
In the slave boson approach the spinon 
is only proportional to the fermion $X^{0 \sigma}$ 
operator in the leading order of $1/N$
expansion \cite{Wang}.

It is also important to note that our diagram technique contains ghost fields 
that cancel infinities to order $O(1/N)$ in the boson propagator.
In the slave boson approach in the radial gauge, the ghost 
fields necessary to treat the Jacobian, cancel infinities in $O(1/N^2)$ 
\cite{Read}.

Using our large-$N$ expansion, we have calculated the boson 
propagator and  the  charge-charge and spin-spin 
correlation functions in $O(1)$.   
In sections III and IV we have shown 
that these quantities agree
with those calculated by slave boson \cite{Wang} and $X$-operator 
canonical approach
\cite{Gehlhoff}. 
These results, which are far nontrivial, show the consistency 
of our formulation.

We showed that in the N-infinite limit we have well defined electronic 
excitations with renormalized band due to correlations. 
This means (and we think) that the large-N
expansion, in the form presented in this paper, is a better approximation 
to describe high-$Tc$ cuprates in optimal doping and 
overdoped than in underdoped close to antiferromagnetism.    

We believe that our formulation will be useful 
(and simpler than previous ones) to calculate quantities in
the next order in $O(1/N)$, such as superconductivity and
electronic self-energy corrections.  We are also planning 
to extend the method to the 
case of finite $J$. 




\noindent{\bf Acknowledgments}

The authors thank to O. Zandron for valuable discussions. 
A.G. wishes to acknowledge many interesting discussions with
A. Dobry, O. Foj\'on and R.Zeyher.



\end{document}